\documentclass[aps,prb,preprint,showpacs,amsmath,amssymb]{revtex4-1}	

\usepackage{graphicx}
\usepackage{subfigure}
\usepackage{multirow}
\usepackage[version=3]{mhchem}
\usepackage{longtable}
\usepackage{tabularx}
\usepackage{color}
\usepackage{cases}
\usepackage{amsmath}
\usepackage{mathtools}
\usepackage{braket}
\usepackage{longtable}
\usepackage{float}
\usepackage{xr}
\usepackage{bm}

\begin{document}

\title[Quantum-Accurate SNAP models for Ni-Mo Binary Alloys and FCC Metals]{Quantum-Accurate Spectral Neighbor Analysis Potential Models for Ni-Mo Binary Alloys and FCC Metals}
\author{Xiang-Guo Li, Chongze Hu, Chi Chen, Zhi Deng, Jian Luo and Shyue Ping Ong}\email{ongsp@eng.ucsd.edu}
\affiliation{Department of NanoEngineering, University of California San Diego, 9500 Gilman Dr, Mail Code 0448, La Jolla, CA 92093-0448, United States}

\begin{abstract}
In recent years, efficient inter-atomic potentials approaching the accuracy of density functional theory (DFT) calculations have been developed using rigorous atomic descriptors satisfying strict invariances, for example, to translation, rotation, permutation of homonuclear atoms, among others. In this work, we generalize the spectral neighbor analysis potential (SNAP) model to bcc-fcc binary alloy systems. We demonstrate that machine-learned SNAP models can yield significant improvements even over well-established, high-performing embedded atom method (EAM) and modified EAM (MEAM) potentials for fcc Cu and Ni. We also report on the development of a SNAP model for the fcc Ni-bcc Mo binary system by machine learning a carefully-constructed large computed data set of elemental and intermetallic compounds. We demonstrate that this binary Ni-Mo SNAP model can achieve excellent agreement with experiments in the prediction of Ni-Mo phase diagram as well as near-DFT accuracy in the prediction of many key properties such as elastic constants, formation energies, melting points, etc., across the entire binary composition range. In contrast, the existing Ni-Mo EAM has significant errors in the prediction of the phase diagram and completely fails in binary compounds. This work provides a systematic model development process for multicomponent alloy systems, including an efficient procedure to optimize the hyper-parameters in the model fitting, and paves the way to long-time, large-scale simulations of such systems.
\end{abstract}

\maketitle

\section{Introduction}

Machine learning (ML) models based on robust local environment descriptors have recently emerged as an approach to describe the potential energy surface (PES) of systems of atoms with near-quantum accuracy at several of orders magnitude lower cost than \textit{ab initio} methods.\cite{Behler2007,Bartok2010,Rupp2011,Bartok2013,Thompson2015,Chmiela2016} Effective local environment descriptors must be invariant under translation, rotation, and permutation of homonuclear atoms, and have the properties of uniqueness and differentiability.\cite{Huan2017} Examples of such descriptors include symmetry functions\cite{Behler2014,Behler2007}, smooth overlap of atomic positions (SOAP)\cite{Szlachta2014,Bartok2013}, bispectrum\cite{Thompson2015,Bartok2010}, Coulomb matrix\cite{Faber2015,Rupp2015,Rupp2011}, among others. A typical approach is to fit the PES as a function of these descriptors by machine learning on \textit{ab initio} data sets, using techniques ranging from simple linear regression\cite{Chen2017,Thompson2015} to kernel ridge regression\cite{Huan2017,Chmiela2016} to neural networks\cite{Artrith2017,Smith2017,Faraji2017,Kolb2017}.

Thus far, the development of ML potentials based on local environment descriptors have largely been limited to elements and oxides. The Gaussian approximation potential (GAP) using the SOAP descriptor has been applied on Si\cite{Bartok2013}, C\cite{Deringer2017,Rowe2018}, W\cite{Szlachta2014}, P\cite{C8FD00034D}, and Fe\cite{Dragoni2017}, and neural network models based on symmetry functions have been fitted for Si\cite{Behler2008}, C\cite{Khaliullin2010}, Na\cite{Eshet2012}, ZnO\cite{Artrith2011}, \ce{TiO2}\cite{Artrith2016}, GeTe\cite{Sosso2012}, and \ce{Li3PO4}\cite{Li2017}. Thompson and Wood\cite{Thompson2015,Wood2017} have developed linear and quadratic models based on the SO(4) bispectrum - the Spectral Analysis Neighbor Potential or SNAP - for bcc Ta and W. \citet{Chen2017} later showed that a linear SNAP model can achieve near-DFT accuracy across a wide range of properties and outperforms embedded atom method (EAM) and modified EAM (MEAM) in the bcc Mo system. Only recently, neural network models utilizing the symmetry function descriptors have been extended to Al-Mg-Si\cite{Kobayashi2017} and \ce{Li_$x$Si}\cite{Artrith2018} alloy systems. The extension of ML models to multi-component oxides and alloys generally leads to a large expansion in the size of the descriptor feature vector, and correspondingly, an explosion in the quantity of data (and hence computational cost) necessary for model fitting.

In this work, we will apply the linear SNAP approach to the bcc Mo-fcc Ni binary alloy system as well as present an investigation of its performance on fcc metals (Cu and Ni). Our choice of model is motivated by the relatively simple functional form of the linear SNAP approach, which reduces the computational effort for model training and minimizes the risk of over-fitting. While \citet{Wood2017} have recently shown a quadratic SNAP model can achieve higher accuracies, this improvement comes at a large increase in the number of fitted coefficients (e.g., 481 for quadratic SNAP vs 31 for linear SNAP in Ta) and consequently, a large increase in the training data set required, an issue which is severely exacerbated in multi-component systems. On the other hand, the efficiency of the Gaussian approximation potential based on the SOAP descriptor\cite{Szlachta2014} depend on the size of the underlying reference set, which would again be greatly compounded in a multi-component system. Ni-Mo alloys are of immense technological interest due to their high corrosion resistance, low thermal-expansion coefficients, hardness and catalytic properties\cite{Krstajic2008,Donten2005,Han2010,Hu2017}. The currently available Ni-Mo embedded atom method (EAM) force field cannot provide satisfactory accuracy on many properties, and even fails in binary compounds. We demonstrate that the ML SNAP models for both fcc and fcc-bcc mixed binary systems can achieve near-quantum accuracy across a wide range of properties, including energies, forces, elastic properties, melting points, surface energy, etc., consistently outperforming the EAM models especially in the binary systems.

\section{Methods}\label{sec:method}

\subsection{Bispectrum and SNAP formalism}

The bispectrum and SNAP formalism have been extensively covered in previous works\cite{Thompson2015,Bartok2010}. We will only briefly describe the key concepts here for completeness. 

The atomic environment is described by the neighbor density $\rho_i(\textbf{r})$ for each atom $i$ at coordinates $\textbf{r}$ , defined as follows:
\begin{equation}
\rho_i(\textbf{r}) = \delta(\textbf{r}) +\\ \sum_{r_{ij}<R_c}f_c(r_{ij})w_{atom}^j\delta(\textbf{r}-\textbf{r}_{ij}).\label{eq1}
\end{equation}

where $\delta(\textbf{r}-\textbf{r}_{ij})$ is the Dirac delta function centered at each neighboring site, the cutoff function $f_c$ ensures a smooth decay for the neighbor atomic density to zero at the cutoff radius $R_c$, and the dimensionless atomic weights $w_{atom}^j$ distinguish different atom types. This density function can then be expanded in 4D hyper-spherical harmonics $U_{m,m^{'}}^j(\theta,\phi,\theta_0)$, as
\begin{equation}
\rho_i(\textbf{r}) = \sum_{j=0}^\infty \sum_{m,m^{'} = -j}^ju_{m,m^{'}}^jU_{m,m^{'}}^j(\theta,\phi,\theta_0) .\label{eq2}
\end{equation}

where the radial component is converted into a third polar angle defined by $\theta_0 = \theta_0^{max}\frac{r}{R_c}$, $\theta_0^{max}$ is the angle conversion function, which was kept at the default value of 0.99363$\pi$ in this work, and the coefficients $u_{m,m^{'}}^j$ are given by the inner product $\braket{U_{m,m^{'}}^j|\rho}$. The bispectrum coefficients are then given by:
\begin{eqnarray}
\textbf{\it B}_{j_1,j_2,j} = \sum_{m_1,m_1^{'} = -j_1}^{j_1}\sum_{m_2,m_2^{'} = -j_2}^{j_2}\sum_{m,m^{'} = -j}^j(u_{m,m^{'}}^j)^*\cdot C_{j_1m_1j_2m_2}^{jm}\times C_{j_1m_1^{'}j_2m_2^{'}}^{jm^{'}} u_{m_1^{'},m_1}^{j_1}u_{m_2^{'},m_2}^{j_2},\label{eq3}
\end{eqnarray}
where $C_{j_1m_1j_2m_2}^{jm}$ are Clebsch-Gordon coefficients. In practice, $j$, $j_1$, and $j_2$ need to be truncated with $j, j_1, j_2 \leq j_{max}$. We found that an order of three for the bispectrum coefficients ($j_{max} = 3$) is sufficient based on our tests, consistent with previous works,\cite{Bartok2010,Thompson2015,Chen2017} which gives a total of 31 projected bispectrum components.

In the SNAP formalism, the total energy $E_{SNAP}$ and forces $F_{SNAP}$ are expressed as a linear function of the 31 projected bispectrum components $B_k$ ($k = \{j, j_1, j_2\}$) and their derivatives, as follows:
\begin{equation}
\begin{aligned}
E_{SNAP} & =  \sum_{i=1}^N\beta_0^{\alpha_i}+\sum_{i=1}^N\sum_{k=\{j, j_1, j_2\}}\beta_k^{\alpha_i}B_k^i\\
\textbf{\textit{F}}_{SNAP}^j & =  -\sum_{i=1}^N{\pmb{\beta}}^{\alpha_i}\pmb{\cdot} \frac{\partial \textbf{\textit{B}}^i}{\partial \textbf{r}_j}\label{eq4}
\end{aligned}
\end{equation}
where $\beta_k^{\alpha_i}$ are the fitting parameters in the linear model, $\alpha_i$ specifies the atom type of atom $i$.

The calculations of bispectrum coefficients (the features) for all the training structures were performed using the implementation in the LAMMPS software\cite{Plimpton1995} by \citet{Thompson2015}. The cutoff radius $R_c$ and atomic weight $w_{atom}$ were treated as hyperparameters fitted during the training of the model, as outlined in subsequent sections.

\subsection{SNAP model fitting}

\begin{figure*}[htp]
\includegraphics[width=1.0 \textwidth]{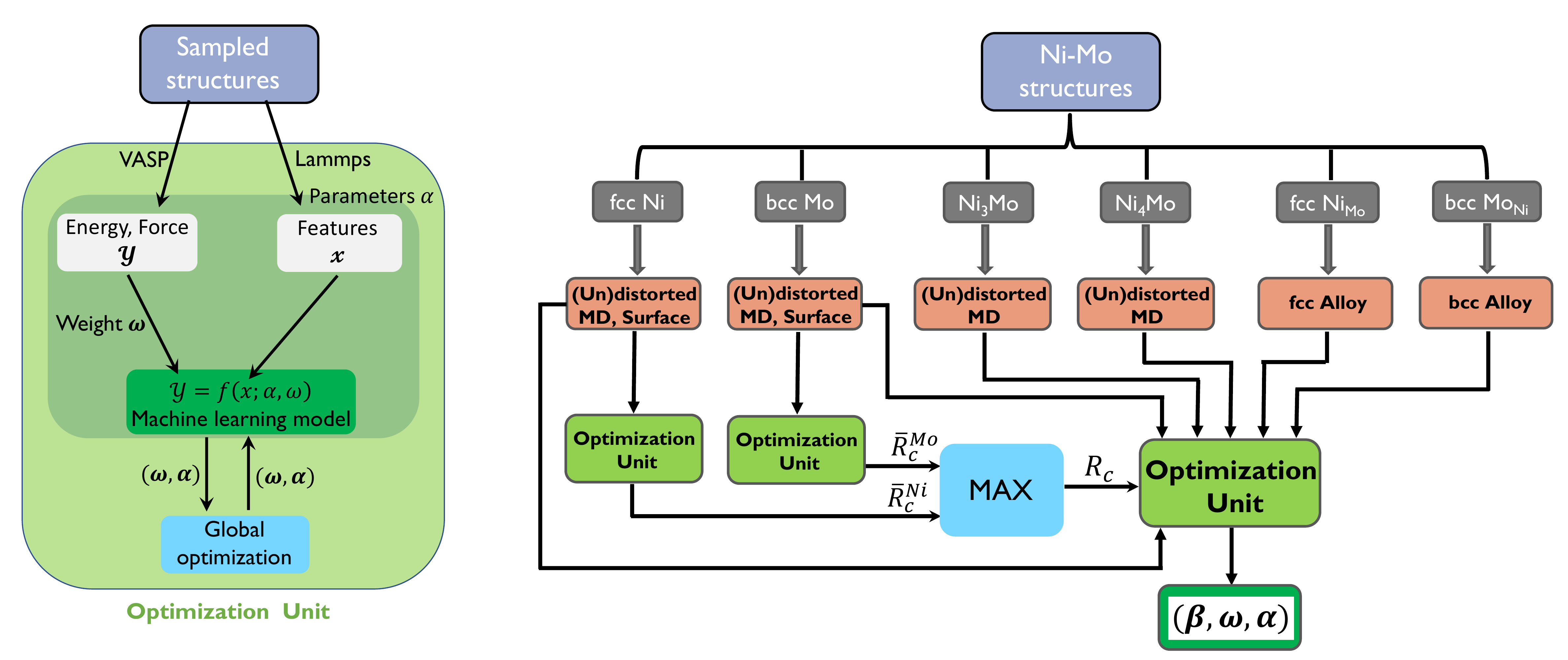}
\caption{\label{fig:model}Fitting workflow for binary alloy SNAP model. Left panel shows one optimization unit developed by \citet{Chen2017}, which optimizes both the model parameters and hyperparameters with respect to DFT calculated energies, forces, and elastic constants. Right panel shows the workflow for binary alloy system. $\alpha$ denotes the parameters (hyperparameters) for the bispectrum calculations, while {\pmb{$\beta$}} denotes the model parameters.}
\end{figure*}

For elemental fcc systems, we adopted the potential fitting workflow developed by \citet{Chen2017}, as shown in the left panel of figure \ref{fig:model}. We denote this whole optimization process as one optimization unit, which consists of two optimization loops. The inner loop optimizes the ML model parameters by mapping the descriptors (bispectrum coefficients) to DFT energies and forces. The outer loop optimizes the hyper-parameters by minimizing the difference between the model predicted material properties, i.e. elastic tensors, and DFT computed values. As introduced by \citet{Chen2017}, the hyperparameters are the data weights ($\omega$) from different data groups, and the parameters ($\alpha$) used in bispectrum calculations, i.e. the radius cutoff $R_c$ and atomic weight $w_{atom}$. In elemental system, the atomic weight can be set as unity. The inner loop fitting of the model coefficients was performed with the least squares algorithm implemented in the scikit-learn package.\cite{pedregosa:hal-00650905} The outer loop optimization was done using the differential evolution algorithm\cite{Neri2010} from the SciPy package\cite{jones2014scipy}.

For the binary Ni-Mo alloy system, there are four parameters ($R_c^{\mbox{\scriptsize{Ni}}}$, $R_c^{\mbox{\scriptsize{Mo}}}$, $w^{i,\mbox{\scriptsize{Ni}}}_{atom}$, $w^{j,\mbox{\scriptsize{Mo}}}_{atom}$) in the bispectrum calculations, two for each element. As it would be far too time-consuming to optimize all parameters simultaneously, we propose instead a two-step model fitting workflow, as shown in the right panel of Figure \ref{fig:model}. The first step involves the independent optimization of the radius cutoff $R_c$ for each elemental system, i.e. Ni and Mo. The maximum of the two optimized radius cutoffs, $\max(R_c^{\mbox{\scriptsize{Ni}}}, R_c^{\mbox{\scriptsize{Mo}}})$, is then used as a common radius cutoff for the binary Ni-Mo system. The use of a common radius cutoff is to maintain symmetric interactions between neighboring atoms of different types, i.e., the interaction between a Mo and a neighboring Ni should be the same as that between a Ni and a neighboring Mo for the same distance. The atomic weight for the element with larger radius cutoff (Mo in this case) is then set at unity. Therefore, only one parameter, the atomic weight for the other element, needs to be optimized in the second optimization step, as shown in the right panel of Figure \ref{fig:model}.

\subsection{Training data generation} 

A diverse set of the training data encompassing a good range of atomic local environments is critical to developing an effective and robust potential. Our training data can be divided into five categories:
\begin{enumerate}
    \item Undistorted ground state structures for Ni, Mo and the two binary intermetallics \ce{Ni3Mo} and \ce{Ni4Mo};
    \item Distorted structures constructed by applying strains of $-10\%$ to 10$\%$ at 1$\%$ intervals to a bulk supercell in six different modes, as described in the work by \citet{DeJong2015};
    \item Surface structures of elemental structures obtained from the Crystalium database\cite{Tran2016a,crystalium}, which include the surface structures with Miller indices up to three; 
    \item Snapshots from $NVT$ \textit{ab initio} molecular dynamics (AIMD) simulations of the bulk supercell at 300, 1000, and 3000 K at the equilibrium 0K volume. In addition, snapshots were also obtained from $NVT$ AIMD simulations at 300K at 90\% and 110\% of the equilibrium 0K volume. Forty snapshots were extracted from each AIMD simulation at intervals of 0.1 $ps$; 
    \item Alloy structures constructed by partial substitution of supercells of the bulk fcc Ni with Mo and the bulk bcc Mo with Ni. Compositions of the form \ce{Ni_$x$Mo_$1-x$} were generated with $x$ ranging from 0 at\% to 100 at\% at intervals of 12.5 at\%.
\end{enumerate}

The supercells used for the distorted structures and AIMD simulations are $3 \times 3 \times 3$ conventional cell for all elemental systems, $3 \times 3 \times 2$ for \ce{Ni3Mo}, and $2 \times 2 \times 3$ for \ce{Ni4Mo}. The Mo-substituted Ni fcc alloy (Ni$_{\mbox{\scriptsize{Mo}}}$) structures were generated in three steps. First, a $2\times 2\times 2$ supercell of Ni was doped with 1, 2, 3, 4, 5, 9, 13, 17, 21, 25, 29, 30, 31, and 32 Mo atoms, respectively. Second, for each doped structure, we performed a structure enumeration\cite{Hart2008} to generate all symmetrically distinct structures, from which up to 100 random structures are selected. Third, we performed a structure relaxation for each selected structure. Both the unrelaxed and relaxed structures were included in our data set. The Ni-substituted Mo bcc alloy (Mo$_{\mbox{\scriptsize{Ni}}}$) structures were constructed using the same procedure with a $2 \times 2 \times 2$ supercell. In addition, since the bcc conventional cell contains half the number of atoms of the fcc conventional cell, we also generated low-concentration Ni-substituted Mo by doping a $3 \times 3 \times 3$ Mo supercell with $1-4$ Ni atoms. 

\subsection{DFT calculations} 

All DFT calculations were performed using the Perdew-Burke-Ernzerhof (PBE)\cite{Perdew1996} exchange correlation functional as implemented in the Vienna \textit{ab initio} simulation package (VASP)\cite{Kresse1996a} within the projector augmented wave (PAW) approach\cite{Blochl1994a}. The kinetic energy cutoff was set to 520 eV and the $k$-point density was at least 3000 per reciprocal atom. Energies and forces were converged to within 10$^{-5}$ eV and 0.02 eV/\AA, respectively. The AIMD simulations were performed with a single $\Gamma$  $k$ point and were non-spin-polarized. However, the energy and force calculations on the snapshots were performed using the same parameters as the rest of the data. All structure manipulations and analysis of DFT computations were carried out using the Python Materials Genomics (pymatgen)\cite{Ong2013} library and automation of calculations was carried out using the Fireworks software\cite{Jain2015b}. 

\subsection{Melting points and phase diagram} 

The melting temperatures $T_m$ were calculated using the solid-liquid coexistence approach.\cite{Morris1994} MD simulations were performed using the $30\times 15\times 15$ bcc (13,500 atoms) and $30\times 10\times 10$ (12,000 atoms) fcc supercells under zero pressure at different temperatures. The time step was set to 1 fs, and simulations were carried out for at least 100 ps. The $T_m$ was identified when the initial solid and liquid phases were at equilibrium (no interface motion). 

With the fully equilibrated solid-liquid structures at the melting points, we conducted hybrid MC/MD simulations to calculate the solidus and liquidus lines at different temperatures. At each temperature below $T_m$, the global composition of dopant atoms was adjusted to find solid-liquid equilibrium phases. The solidus and liquidus lines were then determined by calculating the composition of dopant atoms in the solid and liquid phases, respectively. To reduce statistical errors, all calculations were averaged based on five random structures in the last 10 ps.

The CALculation of PHAse Diagrams (CALPHAD) Ni-Mo phase diagram\cite{lukas2007computational} was constructed using the Pandat software.\cite{Chen2003} In the CALPHAD approach, the liquid phase and two solid terminal phases of Ni-Mo alloy were treated using a subregular solution model,\cite{FRISK1990311} and the model parameters were fitted to experimental data on phase equilibria in the Mo-Ni system.

\subsection{Data availability}

To ensure the reproducibility and use of the models developed in this work, all data (structures, energies, forces, etc.) used in model development as well as the final fitted model coefficients have been published in an open repository (https://github.com/materialsvirtuallab/snap). We will also work with the developers of LAMMPS to include the elemental and binary SNAP models in the LAMMPS software package.

\section{Results}

\subsection{Optimized SNAP model coefficients}

The optimized SNAP model coefficients ($\beta_k$ in equation \ref{eq4}) for elemental fcc Ni, Cu and mixed bcc Ni-fcc Mo systems are provided in the Supplementary Information (SI). The optimized elemental cutoff radius $R_c^{\mbox{\scriptsize{Ni}}}$ and $R_c^{\mbox{\scriptsize{Cu}}}$ are 3.9 and 3.7 \AA, respectively, slightly larger than the second nearest neighbor distance in the respective fcc crystals. For bcc Mo, the optimized $R_c^{\mbox{\scriptsize{Mo}}}$ is 4.6 \AA.\cite{Chen2017} For the mixed bcc Ni-fcc Mo model, the overall cutoff radius $R_c$ is set as $\max(R_c^{\mbox{\scriptsize{Ni}}}, R_c^{\mbox{\scriptsize{Mo}}})= 4.6$ \AA, $w_{atom}^{\mbox{\scriptsize{Mo}}}$ = 1.0, and the optimized value for $w_{atom}^{\mbox{\scriptsize{Ni}}}$ is 0.5.

\subsection{Performance of fcc Ni SNAP model}

We will first discuss the performance of the SNAP model for fcc metals, given that the SNAP approach has hitherto been applied to only bcc metals such as W, Ta and Mo. Here, we will focus our discussion on the elemental fcc Ni SNAP model and compare its performance to that for the binary fcc Ni-bcc Mo model. We have constructed a SNAP model for Cu as well using a similar approach, and the qualitative results are similar and reported in the Supplementary Information. 

\subsubsection{Energies and forces.}

\begin{figure*}[t]
\includegraphics[width=0.9 \textwidth]{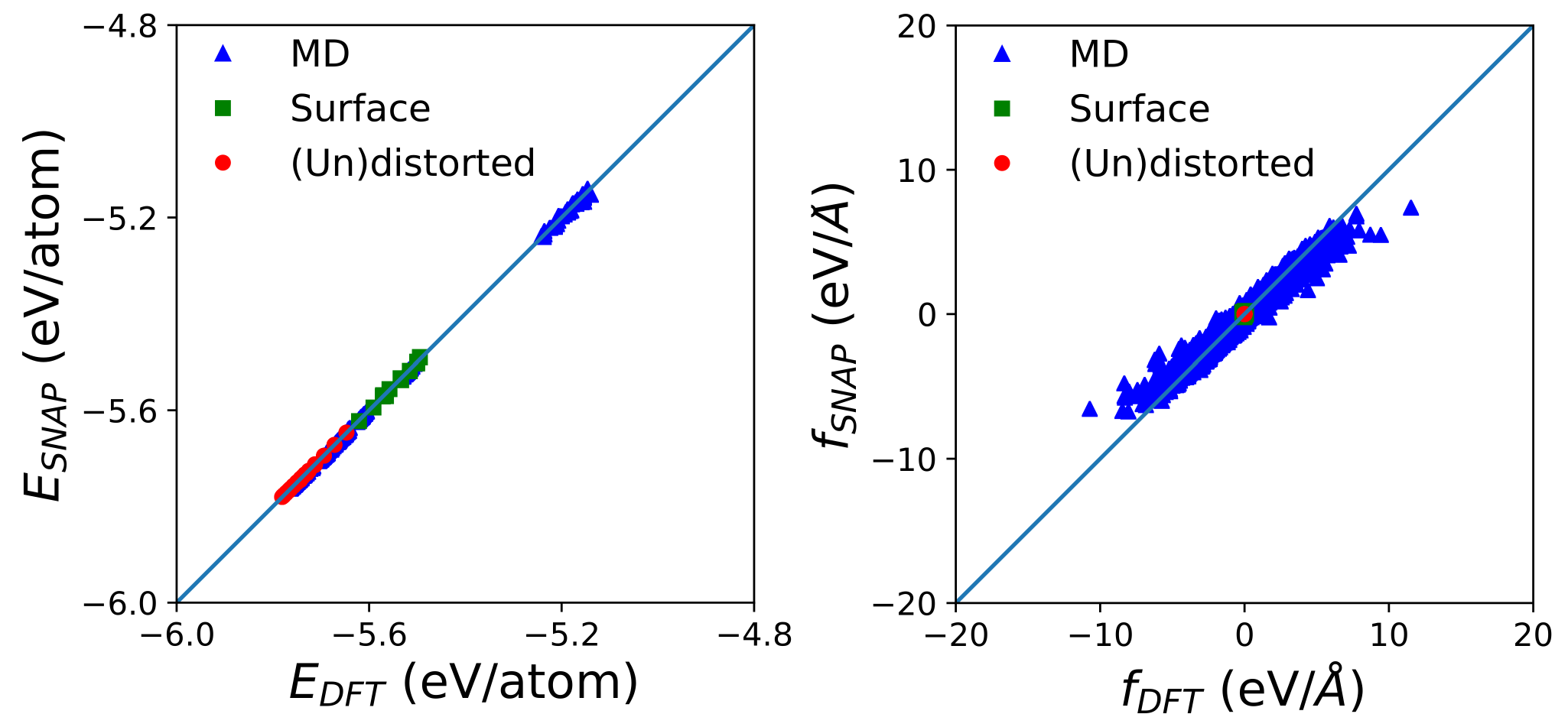}
\caption{\label{fig:e_f}Plot of SNAP predictions versus DFT for energies (left panel) and forces (right panel) in elemental Ni system for different data groups. The MAE for energy and force are 1.2 meV/atom, 0.05 eV/\AA, respectively.}
\end{figure*}

A comparison of the DFT and SNAP predicted energies and forces for elemental Ni is shown in Figure \ref{fig:e_f}. For both energies and forces, SNAP model predictions are in line with the DFT results with a unity slope. EAM potentials are well known to have a good performance in fcc metals\cite{Sheng2011}. The mean absolute error (MAE) in the energies and forces (relative to DFT) for the EAM potential\cite{Zhou2004} are 10.6 meV/atom and 0.06 eV/\AA, respectively, while that for the MEAM potential\cite{Asadi2015} are 17.8 meV/atom and 0.08 eV/\AA, respectively. The Ni SNAP model can achieve a much lower MAE in energy of 1.2 meV/atom, and slightly improved MAE in force of 0.05 eV/\AA. 

To further validate our model, we generated test structures by performing additional Ni surface calculations with Miller indices up to four, and also extracting 40 snapshots from AIMD simulations on the vacancy-containing supercell of Ni at 1000 K. The predicted MAEs for the energies and forces are 2.3 meV/atom and 0.08 eV/\AA, respectively, comparable to the model performance on the training datasets. This validation of the model on previous unseen data indicates that the model can be generalized.
\begin{table*}[htp]
\caption{\label{tab:table1} Comparison of the calculated and experimental melting points ($T_m$), elastic constants ($c_{ij}$), Voigt-Reuss-Hill\cite{Hill1952} bulk modulus (B$_{VRH}$), shear modulus (G$_{VRH}$), Poisson's ratio ($\mu$), vacancy formation energy ($E_v$), and migration energy ($E_m$) of fcc Ni. Error percentages of the SNAP, EAM and MEAM predictions with respect to DFT values are shown in parentheses. The values of B$_{VRH}$, G$_{VRH}$ and $\mu$ in Exp. column are derived from the experimental elastic constants.}
\begin{tabularx}{1.0\textwidth}{cccccc}
\hline\hline
\noalign{\smallskip}
&\quad DFT&\quad SNAP&\quad EAM\cite{Zhou2004}& \quad MEAM\cite{Asadi2015}&\quad Exp.\\
\hline
\noalign{\smallskip}
$T_m$ (K)&\quad $-$ &\quad  1785 &\quad1520&\quad 1765&\quad1728 \\
$c_{11} $ (GPa)&\quad 276 &\quad 276 ($0.0\%$)&\quad 248 ($-10.1\%$)&\quad 260 ($-5.8\%$)&\quad 261\cite{Alers1960} \\
$c_{12} $ (GPa)&\quad 159 &\quad 159 ($0.0\%$)&\quad147 ($-7.5\%$)&\quad 151 ($-7.5\%$)&\quad 151\cite{Alers1960} \\
$c_{44} $ (GPa)&\quad 132 &\quad 132 ($0.0\%$)&\quad  125 ($-5.3\%$)&\quad 131 ($-0.8\%$)&\quad 132\cite{Alers1960} \\
B$_{VRH}$ (GPa)&\quad 198 &\quad198 ($0.0\%$)&\quad 181 ($-8.6\%$)&\quad 187 ($-5.6\%$)&\quad $188$ \\
G$_{VRH}$ (GPa)&\quad  95&\quad 95 ($0.0\%$)&\quad  87 ($-8.4\%$)&\quad 92 ($-3.2\%$)&\quad $93$ \\
$\mu$&\quad 0.29 &\quad 0.29 ($0.0\%$)&\quad  0.29 ($0.0\%$)&\quad 0.29 ($0.0\%$)&\quad  $0.29$ \\
$E_v$ (eV)&\quad 1.46 &\quad 1.68 ($15.1\%$)&\quad 1.68 ($15.1\%$)&\quad 1.16 ($-20.5\%$)&\quad $1.54-1.80$\cite{Megchiche2006} \\
$E_m$ (eV)&\quad 1.12 &\quad 1.07 ($-4.5\%$)&\quad 0.90 ($-19.6\%$)&\quad 1.46 ($30.4\%$)&\quad  $1.01-1.48$\cite{Megchiche2006} \\
$E_a = E_v + E_m$ (eV)&\quad 2.58 &\quad  2.75 ($6.6\%$)&\quad  2.58 ($0\%$)&\quad 2.62 ($1.6\%$)&\quad  $2.77-2.95$\cite{Megchiche2006} \\

\hline\hline
\end{tabularx}
\end{table*} 

\begin{figure}[htb]
\includegraphics[width= 0.7\textwidth]{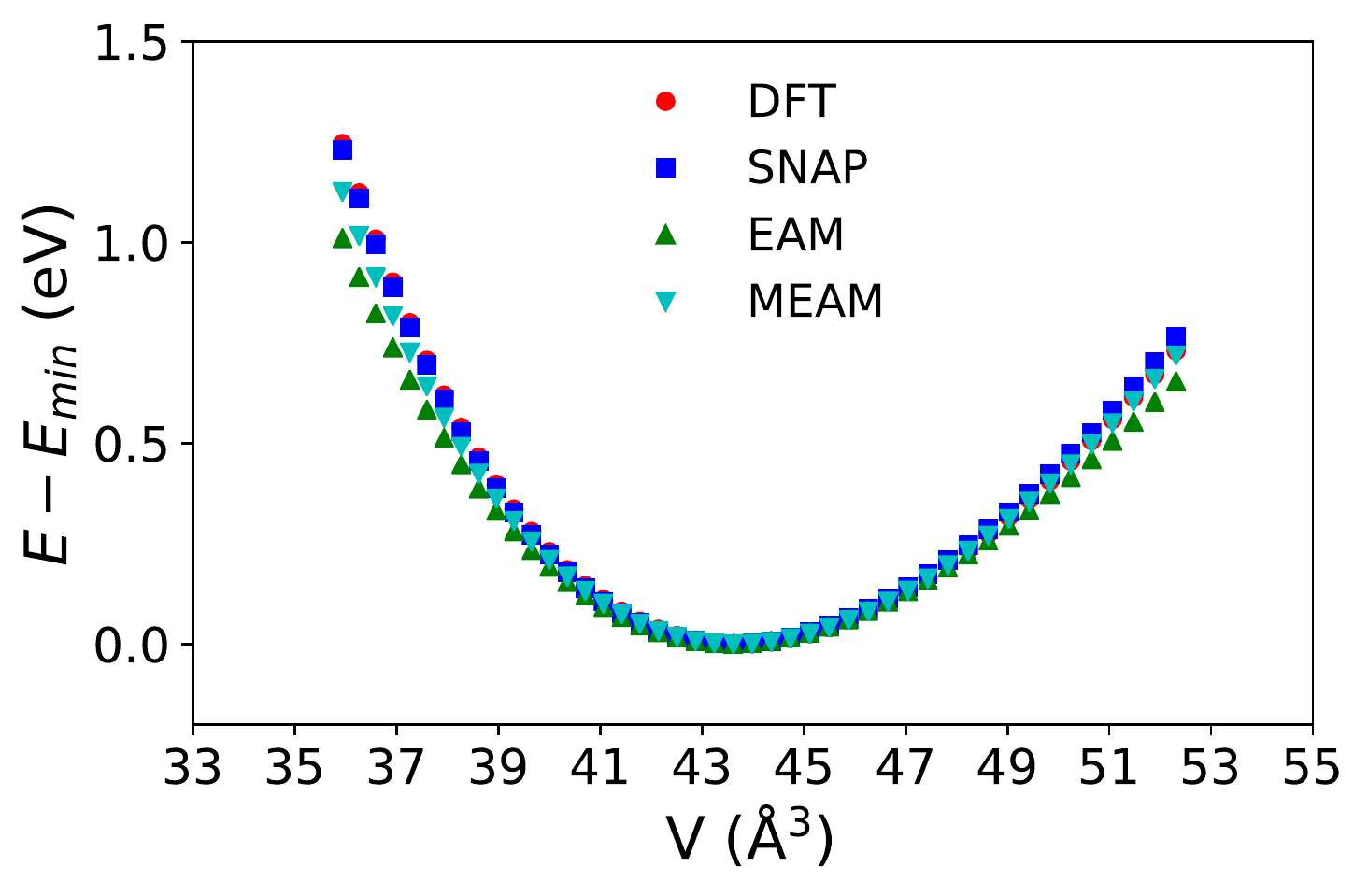}
\caption{\label{fig:eof} Energy vs volume curves of a conventional fcc Ni cell for the DFT, SNAP, EAM, and MEAM models. The energy at the equilibrium volume has been set as the zero reference.}
\end{figure}

\begin{figure}[htb]
\includegraphics[width= 0.7\textwidth]{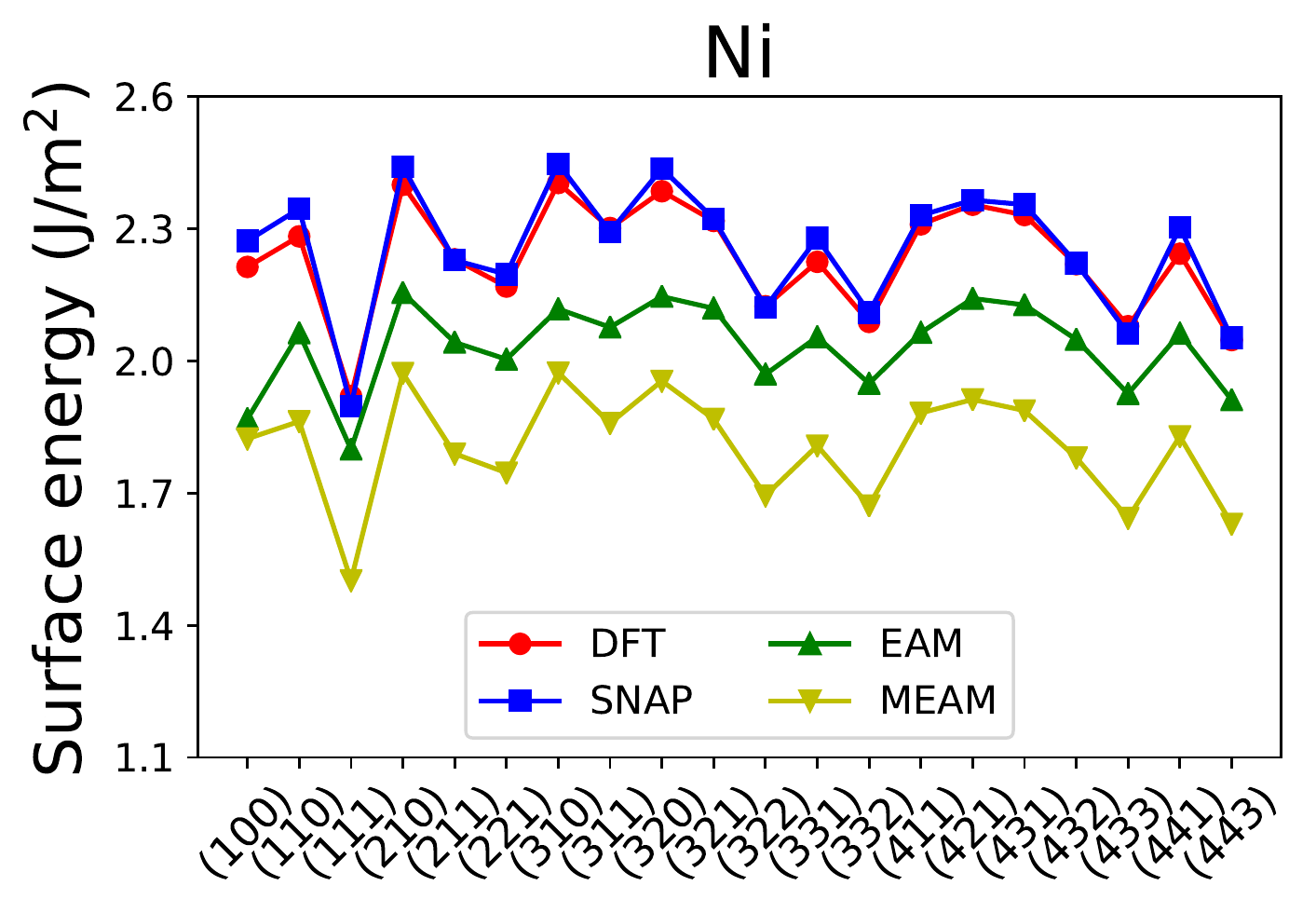}
\caption{\label{fig:sur} Comparison of calculated surface energies for Ni surfaces with Miller indices up to a maximum of four using DFT, SNAP, EAM, and MEAM.}
\end{figure}

\begin{table*}[htb]
\caption{\label{tab:table2} Comparison of the MAEs in predicted energies and forces relative to DFT for the three SNAP models (elemental Ni, elemental Mo\cite{Chen2017}, and binary Ni-Mo) and the binary Ni-Mo EAM model\cite{Zhou2004}. The ``Overall'' column refers to the MAE across the entire training dataset.}
\begin{tabularx}{\textwidth}{ccccccccc}
\hline\hline
\noalign{\smallskip}
&\quad Model&\quad Mo&\quad \ce{Ni4Mo}&\quad \ce{Ni3Mo}&\quad Mo$_{\mbox{\scriptsize{Ni}}}$&\quad Ni$_{\mbox{\scriptsize{Mo}}}$&\quad Ni&\quad Overall\\
\hline
\noalign{\smallskip}
\multirow{ 4}{*} {Energy (meV/atom)}&\quad Ni SNAP &\quad $-$ &\quad$-$ &\quad$-$&\quad$-$&\quad$-$&\quad 1.2&\quad$-$\\
&\quad Mo SNAP &\quad 13.2 &\quad$-$ &\quad$-$&\quad$-$&\quad$-$&\quad $-$&\quad$-$\\
&\quad Ni-Mo SNAP &\quad 16.2 &\quad 4.0 &\quad 5.2&\quad22.7&\quad33.9&\quad 7.9&\quad22.5\\
&\quad EAM &\quad58.9 &\quad 211.2&\quad255.6&\quad46.5&\quad147.6&\quad 10.6&\quad117.2\\
\noalign{\smallskip}
\hline
\noalign{\smallskip}
\multirow{ 4}{*} {Force (eV/\AA)}&\quad Ni SNAP &\quad $-$ &\quad$-$ &\quad$-$&\quad$-$&\quad$-$&\quad 0.05&\quad$-$\\
&\quad Mo SNAP &\quad 0.25 &\quad$-$ &\quad$-$&\quad$-$&\quad$-$&\quad $-$&\quad$-$\\
&\quad Ni-Mo SNAP &\quad 0.29 &\quad 0.14&\quad0.16&\quad0.13&\quad0.55&\quad 0.11&\quad0.23\\
&\quad EAM &\quad 0.31 &\quad 0.20&\quad0.19&\quad0.21&\quad0.57&\quad 0.06&\quad 0.26\\
\hline\hline
\end{tabularx}
\end{table*}

\subsubsection{Materials properties}

Table \ref{tab:table1} provides a comparison of the Ni SNAP model predictions of the melting points and elastic properties with DFT, EAM/MEAM potentials and experiments\cite{Alers1960}. We find that both the SNAP and MEAM models predict melting points that are in excellent agreement (within 2-3\%) with the experimental value, but the EAM model greatly underestimates the melting point by $\sim 12\%$. The Ni elastic moduli predicted by the SNAP model are in extremely good agreement with the DFT, but those predicted by the MEAM model are much closer to the experimental values. These differences are the result of the data used for model fitting - the SNAP model was fitted using DFT-calculated data, while the MEAM model\cite{Asadi2015} was fitted using the experimental elastic moduli. The EAM-predicted elastic moduli deviates significantly from both the DFT and experimental values. The SNAP model also predicts vacancy formation and migration energies\cite{Henkelman2000} that are much closer to the DFT values. The EAM model greatly underestimates $E_m$ by $\sim20\%$, while MEAM significantly overestimates $E_m$ by more than $30\%$.

Figure \ref{fig:eof} shows the equation of state curves constructed using the DFT, SNAP, EAM, and MEAM models. We observe that the SNAP curve overlaps with DFT for the whole covered region with volume changes in the range of $-17\%$ to $21\%$ from the equilibrium volume. The EAM potential deviates significantly from the DFT curve at both tensile and compressive strains, and the MEAM potential slightly underestimates the energy at large compressive strains. By fitting the Murnaghan equation of state, the estimated bulk moduli from Fig \ref{fig:eof} are 188, 190, 160, and 177 GPa for DFT, SNAP, EAM, and MEAM, respectively. All three models (SNAP, EAM and MEAM) lead to very similar phonon dispersion curves that are slightly underestimated relative to the DFT curves (see Figure S3 in the Supplementary Information). 

Figure \ref{fig:sur} compares the performance of the Ni SNAP model with DFT\cite{Tran2016a} and the EAM/MEAM models in the prediction of Ni surface energies up to a maximum Miller index of four. The surface energies computed by the SNAP model are in excellent agreement with the DFT calculations, while both the EAM and MEAM models significantly underestimate surface energies. It should be noted that the surfaces with Miller indices beyond three, e.g., (411), (421), etc., are not part of the training dataset and constitutes test data that further validates the applicability of the SNAP model beyond already-seen data.

\subsection{Performance of binary Ni-Mo SNAP model}

In this section, we will discuss the performance of the optimized binary Ni-Mo SNAP model. We will not only compare the performance of the binary SNAP model to DFT and EAM, but also the performance of the binary SNAP model relative to the optimized elemental Ni and Mo SNAP models and discuss any compromises in the performance on the elemental end members in going from a single component to a binary model.

\subsubsection{Energies and forces}

Table \ref{tab:table2} compares the MAEs in predicted energies and forces relative to DFT for the elemental and binary Ni-Mo SNAP models and the binary Ni-Mo EAM model\cite{Zhou2004}. It should be noted that the binary EAM model was constructed from normalized elemental EAM potentials with a relative scaling factor between elements. The relative scaling factor along with the EAM parameters are fitted to the experimentally measured properties, such as lattice constants, elastic constants, vacancy formation energies, heats of solution, etc. As such, our discussion of the relative performance of the Ni-Mo SNAP and EAM models will focus on qualitative trends (especially in the binary alloys and intermetallics) rather than quantitative comparisons.

We may observe that the binary Ni-Mo SNAP model significantly outperforms the binary Ni-Mo EAM model across almost all data sets, with the exception of a larger MAE in predicted forces for pure Ni. In particular, the MAEs in the predicted energies for the binary phases (\ce{Ni4Mo}, \ce{Ni3Mo} and the Mo-doped fcc Ni) are especially large for the EAM model relative to the end member elemental phases, while those for the binary Ni-Mo SNAP are much smaller and comparable for both binary as well as elemental phases. This indicates that a clear bias for the elemental phases in the construction of the binary EAM potential. However, relative to the elemental Mo and Ni SNAP models, the binary Ni-Mo SNAP model clearly sacrifices accuracy on the end member elements with somewhat larger errors in predicted energies and forces for both bcc Mo and fcc Ni. We attribute this decrease in accuracy to the substantially more complex and diverse training structures when fitting binary potential compared with elemental potential. 

\subsubsection{Materials Properties}

Table \ref{tab:table3} compares the elastic properties computed by the elemental and binary SNAP models, the EAM model, DFT and experiments\cite{Alers1960,simmons1971single}. Again, we observe that the binary Ni-Mo SNAP model generally outperforms the binary EAM model in the prediction of the elastic constants, bulk and shear moduli, and Poisson's ratio for the binary intermetallics \ce{Ni3Mo} and \ce{Ni4Mo}. The binary EAM model performs especially poorly in this regard, with absolute percentage errors exceeding 100\% in some instances (e.g., shear modulus and Poisson's ratio for \ce{Ni4Mo}). Compared to the elemental SNAP models, the binary Ni-Mo SNAP model does suffer a slight decrease in prediction accuracy, but still manages to retain better agreement with DFT compared with EAM.

\LTcapwidth=\linewidth
\begin{center}
\begin{longtable}{ccccccc}
\caption{\label{tab:table3} Comparison of elastic constants ($c_{ij}$), Voigt-Reuss-Hill\cite{Hill1952} bulk modulus (B$_{VRH}$), shear modulus (G$_{VRH}$), and Poisson's ratio ($\mu$) for fcc Ni, bcc Mo, and binary compound \ce{Ni4Mo} and \ce{Ni3Mo}. Error percentages of the SNAP (elemental Ni SNAP, Mo SNAP and binary Ni-Mo SNAP) and EAM predictions relative to DFT values are shown in parentheses. The values of B$_{VRH}$, G$_{VRH}$ and $\mu$ in Exp. column are derived from the experimental elastic constants.}\\
\hline\hline
&\quad DFT&\quad Mo SNAP\cite{Chen2017}&\quad Ni SNAP&\quad Ni-Mo SNAP&\quad EAM&\quad Exp.\\
\hline
\textbf{Mo}&&&&&&\\
$c_{11} $ (GPa)&\quad 472 &\quad473 ($0.2\%$)&\quad$-$&\quad 475 ($0.6\%$) &\quad 457 ($-3.2\%$)&\quad 479\cite{simmons1971single} \\
$c_{12} $ (GPa)&\quad 158 &\quad152 ($-3.8\%$)&\quad$-$&\quad163 ($3.2\%$)&\quad 168 ($6.3\%$)&\quad 165\cite{simmons1971single} \\
$c_{44} $ (GPa)&\quad 106 &\quad107 ($0.9\%$)&\quad$-$&\quad 111($4.7\%$)&\quad 116 ($9.4\%$)&\quad 108\cite{simmons1971single} \\
B$_{VRH} (GPa)$&\quad 263 &\quad259 ($-1.5\%$)&\quad$-$&\quad 267 ($1.5\%$)&\quad 264 ($0.4\%$)&\quad $270$ \\
G$_{VRH}$ (GPa)&\quad 124&\quad126 ($1.6\%$)&\quad$-$&\quad 127 ($2.4\%$)&\quad127 ($2.4\%$)&\quad $125$ \\
$\mu$&\quad 0.30&\quad0.29 ($-3.3\%$)&\quad$-$&\quad 0.29($-3.3\%$)&\quad 0.29 ($-3.3\%$)&\quad $0.30$ \\
\textbf{Ni}&&&&&&\\
$c_{11} $ (GPa)&\quad 276&\quad$-$&\quad276 ($0.0\%$)&\quad 269 ($-2.5\%$)&\quad 248 ($-10.1\%$)&\quad 261\cite{Alers1960} \\
$c_{12} $ (GPa)&\quad 159 &\quad$-$&\quad159 ($0.0\%$)&\quad 150 ($-5.7\%$)&\quad 147 ($-7.5\%$)&\quad 151\cite{Alers1960} \\
$c_{44} $ (GPa)&\quad 132&\quad$-$&\quad 132 ($0.0\%$)&\quad 135 ($2.3\%$)&\quad 125 ($-5.3\%$)&\quad 132\cite{Alers1960} \\
B$_{VRH}$ (GPa)&\quad198&\quad$-$&\quad198 ($0.0\%$) &\quad 190 ($-4.0\%$)&\quad 181 ($-8.6\%$)&\quad $188$ \\
G$_{VRH}$ (GPa)&\quad 95&\quad$-$&\quad95 ($0.0\%$)&\quad 97 ($2.1\%$)&\quad 87 ($-8.4\%$)&\quad $93$ \\
$\mu$&\quad 0.29 &\quad$-$&\quad0.29 ($0.0\%$)&\quad 0.28 ($-3.4\%$)&\quad 0.29 ($0.0\%$)&\quad $0.29$ \\
\textbf{\ce{Ni3Mo}}&&&&&&\\
$c_{11} $ (GPa)&\quad 385 &\quad$-$&\quad$-$&\quad 420 ($9.1\%$)&\quad 195 ($-49.4\%$)&\quad $-$ \\
$c_{12} $ (GPa)&\quad166 &\quad$-$&\quad$-$&\quad 197 ($18.7\%$)&\quad 98 ($-41.0\%$)&\quad $-$ \\
$c_{13} $ (GPa)&\quad145 &\quad$-$&\quad$-$&\quad 162 ($11.7\%$)&\quad 98 ($-32.4\%$)&\quad $-$ \\
$c_{23} $ (GPa)& \quad 131 & \quad$-$& \quad$-$& \quad 145 ($10.7\%$)& \quad 107 ($-18.3\%$)& \quad $-$ \\
$c_{22} $ (GPa)& \quad 402 & \quad$-$& \quad$-$& \quad 360 ($-10.4\%$)& \quad 351 ($-12.7\%$)& \quad $-$ \\
$c_{33} $ (GPa)& \quad402 & \quad$-$& \quad$-$& \quad 408 ($1.5\%$)& \quad 295 ($-26.6\%$)& \quad $-$ \\
$c_{44} $ (GPa)& \quad 94 & \quad$-$& \quad$-$& \quad 84 ($-10.6\%$)& \quad 36 ($-61.7\%$)& \quad $-$ \\
B$_{VRH}$ (GPa)& \quad 230 & \quad$-$& \quad$-$& \quad 243 ($5.7\%$)& \quad 156 ($-32.2\%$)& \quad $-$ \\
G$_{VRH}$ (GPa)& \quad 89 & \quad$-$& \quad$-$& \quad 100 ($12.4\%$)& \quad 61 ($-31.5\%$)& \quad $-$ \\
$\mu$& \quad 0.33 & \quad$-$& \quad$-$& \quad 0.32 ($-3.0\%$)& \quad 0.33 ($0.0\%$)& \quad $-$ \\
\textbf{\ce{Ni4Mo}}&&&&&&\\
$c_{11} $ (GPa)& \quad300 & \quad$-$& \quad$-$& \quad 283 ($-5.7\%$)& \quad 172 ($-42.7\%$)& \quad $-$ \\
$c_{12} $ (GPa)& \quad 186 & \quad$-$& \quad$-$&  \quad179 ($-3.8\%$)& \quad 158 ($-15.1\%$)& \quad $-$ \\
$c_{23} $ (GPa)& \quad 166 & \quad$-$& \quad$-$& \quad 164 ($-1.2\%$)& \quad 80 ($-51.8\%$)& \quad $-$ \\
$c_{22} $ (GPa)& \quad313 & \quad$-$& \quad$-$& \quad 326 ($4.2\%$)& \quad 158 ($-49.5\%$)& \quad $-$ \\
$c_{44} $ (GPa)& \quad130 & \quad$-$& \quad$-$& \quad 126 ($-3.1\%$)& \quad125 ($-3.8\%$)& \quad $-$ \\
B$_{VRH}$ (GPa)&  \quad223 & \quad$-$& \quad$-$& \quad 220 ($-1.3\%$)& \quad 161 ($-27.8\%$)& \quad $-$ \\
G$_{VRH}$ (GPa)& \quad 91 & \quad$-$& \quad$-$& \quad 95 ($4.4\%$)& \quad $-56$ ($-162\%$)& \quad $-$ \\
$\mu$& \quad 0.33 & \quad$-$& \quad$-$& \quad 0.31 ($-6.1\%$)& \quad 0.70 ($112\%$)& \quad $-$ \\
\hline\hline
\end{longtable} 
\end{center}

\begin{figure*}[htb]
\includegraphics[width= 1.0\textwidth]{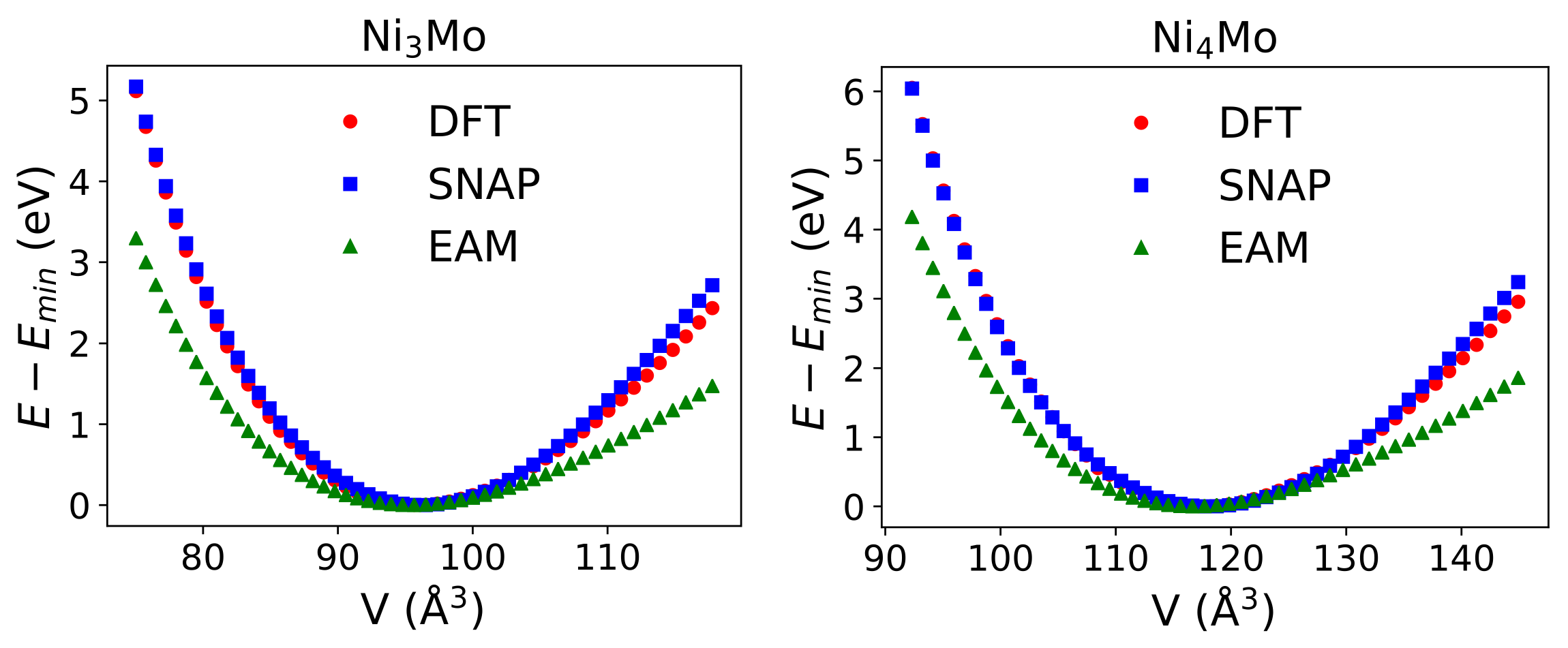}
\caption{\label{fig:NiMoeof} Energy vs volume curves of a conventional \ce{Ni3Mo} (left panel) and \ce{Ni4Mo} (right panel) cell for the DFT, SNAP, and EAM models. The energy at the equilibrium volume has been set as the zero reference.}
\end{figure*}

\begin{table}[htp]
\caption{\label{tab:table4} Melting temperatures (in K) for pure Mo and Ni with different methods. EAM and SNAP values are calculated using the binary force field.}

\begin{tabularx}{1.0\textwidth}{ccccc}
\hline\hline
\noalign{\smallskip}
&\qquad\qquad Experiment&\qquad\qquad CALPHAD&\qquad\qquad EAM&\qquad\qquad SNAP\\
\hline
\noalign{\smallskip}
Pure Mo&\qquad\qquad2898 &\qquad\qquad 2899&\qquad\qquad 3750&\qquad\qquad 3250 \\
Pure Ni&\qquad\qquad 1728&\qquad\qquad 1728&\qquad\qquad 1520&\qquad\qquad 1810\\
\hline\hline
\end{tabularx}
\end{table} 

Figure \ref{fig:NiMoeof} displays the equation of state curves constructed using the DFT, SNAP, and EAM models for the binary compounds \ce{Ni3Mo} and \ce{Ni4Mo}. We observe that for both \ce{Ni3Mo} and \ce{Ni4Mo}, the SNAP curve overlaps with DFT for volume changes in the range of $-21\%$ to $10\%$ from the equilibrium volume, but begins to slightly overestimate the energies with volume expansions beyond $10\%$. The EAM potential completely fails in the equation of state prediction for binary compounds. It significantly underestimates the energies at both tensile and compressive strains. Similar conclusions can be made from the prediction of the phonon dispersion curves - the binary SNAP model produces phonon dispersion curves that are in excellent agreement with DFT for both \ce{Ni3Mo} and \ce{Ni4Mo}, while the EAM potential produces curves with imaginary frequencies, in contradiction to DFT (see Figure S4 in the Supplementary Information).

\begin{figure*}[t]
\includegraphics[width=1.0 \textwidth]{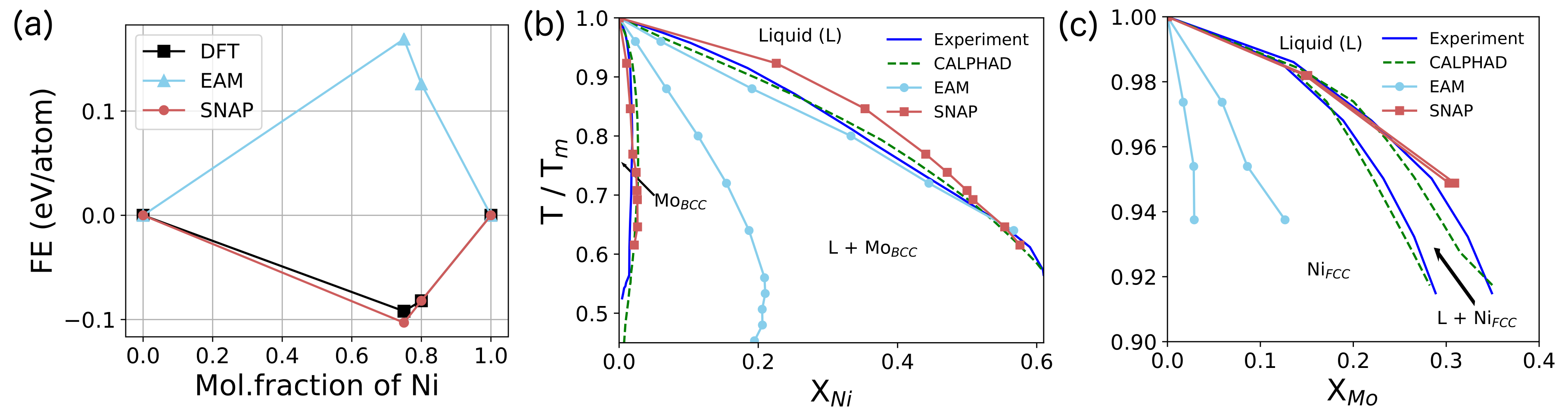}
\caption{\label{fig:phase}Plots of the (a) 0K Ni-Mo pseudo-binary formation energy diagram calculated using DFT, SNAP and EAM, and high-temperature Ni-Mo phase diagram normalized by the melting temperature for (b) Mo-rich domain and (c) Ni-rich domain, from experiments\cite{CASSELTON1964212}, CALPHAD, SNAP and EAM models.}
\end{figure*}

Figure \ref{fig:phase}a compares the 0K Ni-Mo pseudo-binary formation energy diagram calculated using DFT and the binary SNAP and EAM models. The binary EAM model fails to reproduce even qualitatively the convex hull, predicting positive formation energies for \ce{Ni3Mo} and \ce{Ni4Mo}, while the binary SNAP model predictions are in good agreement with DFT. This is consistent with the large prediction error in the energies of the binary intermetallics for the EAM model discussed in the previous section.

Figure \ref{fig:phase}b, c compares the high-temperature ($>$ 1000 K) Ni-Mo phase diagram normalized by the melting temperature calculated from hybrid MC-MD simulations using the binary SNAP and EAM models with those from experiments and CALPHAD.\cite{CASSELTON1964212} Again, we note that the EAM calculated phase diagram exhibits large errors, greatly overestimating the solubility of Ni in Mo by more than 10 times and the melting point of Mo by about 29.4\% (see Table \ref{tab:table4}). In contrast, the SNAP model predicts a maximum solubility of Ni in Mo of about 2.6$\%$ for SNAP, which is in excellent agreement with the experimental value of 1.9\%, and the predicted melting points for Mo are also closer to the experimental values ($~$12.2\% higher, see Table \ref{tab:table4}). The liquidus line calculated by SNAP exhibits concave-like transitions with temperatures, consistent with the experimental phase diagram, but EAM gives a linear relationship. At the Ni-rich domain, the experimental and CALPHAD liquidus and solidus lines are almost overlapping with each other close to $T_m$ of Ni, and this behavior is successfully reproduced by the SNAP model. EAM, on the other hand, shows a large segregation to liquid phases as the temperature decreases from $T_m$, contradictory to experiment. In addition, the solubility of Mo in Ni predicted by EAM is only about one tenth of the experimental value. The main major discrepancy for the binary SNAP model is in the separation of the solidus and liquidus lines. The binary SNAP model predicts an extremely small separation between the solidus and liquidus lines as the temperature is decreased from the Ni melting point, whereas experimentally, these lines are separated by $\sim 5\%$ Ni.

\section{Discussion and Conclusion}

To conclude, we have developed SNAP models for fcc Ni, Cu as well as the binary Ni-Mo system. 

For fcc metals such as Ni and Cu, we find that the elemental SNAP models offer only a modest improvement over well-established EAM/MEAM potentials. This is unlike the case for bcc metals such as Mo, Ta and W, for which EAM/MEAM potentials generally perform relatively poorly and SNAP models have been demonstrated to lead to significant reductions in prediction error in energies, forces and various materials properties\cite{Thompson2015,Wood2017,Chen2017}. 

Where the SNAP formalism truly shines is its extensibility to multi-component systems, achieving consistently low and comparable MAEs in the energies and forces for the elemental end members as well as the binary intermetallics and solid solutions for the bcc Mo-fcc Ni binary alloy system. This performance is achieved using the same simple linear model with a doubling of the number of fitted coefficients and hyper-parameters. We have proposed a two-step fitting approach to efficiently determine the hyper-parameters. In contrast, the EAM model is significantly biased for better error performance in the elemental end members, with extremely large errors and failing even on a qualitative level for the binary intermetallics and alloys. We have successfully applied this SNAP model to reproduce the high-temperature Ni-Mo phase diagram, to excellent agreement with experiments. We believe SNAP models developed using the same principles and approach can enable high accuracy studies of micro-structure and other phenomena requiring  large-scale simulations over long-time scales on multi-component systems.

The main trade-off is the $2-3$ orders of magnitude higher computational cost of SNAP models compared to EAM. Nevertheless, it should be noted that SNAP models still scale linearly with the number of atoms and are orders of magnitude cheaper than DFT calculations. The combination of near-DFT accuracy at several orders of magnitude lower computational cost has enabled us to construct from first principles the high-temperature Ni-Mo phase diagram in Figure \ref{fig:phase}b, which is shown to be in excellent agreement with the experimental phase diagram. This effort, which requires long-time scale simulations of large MD simulation boxes exceeding 10,000 atoms, is beyond the scope of DFT calculations today. Most critically, the binary SNAP model is able to reproduce the correct formation energies and solubilities across a wide range of Ni-Mo structures (fcc, bcc, solid solutions, intermetallics, surfaces), which is indicative of its general applicability to the study of micro-structure and segregation phenomena in this highly important alloy system.

Finally, it is our belief that the development of potential models should account on a holistic basis the trade-offs between prediction accuracy in energies, forces and various properties, computational cost of the models, training data size and extensibility beyond single-component systems.

\centerline {\textbf{Acknowledgements}}

This work was primarily supported by the Office of Naval Research (ONR) Young Investigator Program (YIP) under Award N00014-16-1-2621. The CALPHAD and MC/MD investigations are supported by the Vannevar Bush Faculty Fellowship program sponsored by the Basic Research Office of the Assistant Secretary of Defense for Research and Engineering and funded by the Office of Naval Research through grant N00014-16-12569. The authors also acknowledge computational resources provided by Triton Shared Computing Cluster (TSCC) at the University of California, San Diego, the National Energy Research Scientific Computing Center (NERSC), and the Extreme Science and Engineering Discovery Environment (XSEDE) supported by National Science Foundation under grant number ACI-1053575.

%\bibliography{Ni-Mo}
%

\end{document}